\documentclass{aa}
\usepackage{graphicx}
\usepackage{txfonts}
\usepackage{natbib}
\bibpunct{(}{)}{;}{a}{}{,}

\begin{document}

\title{Two new Intermediate Polars with a soft X-ray component~\thanks{Based on observations
obtained with XMM-Newton, an ESA science mission with instruments and contributions directly
funded by ESA Member States and NASA, and with the Observatoire de Haute Provence (CNRS), France.}}

\author{G.~Anzolin\inst{1,2}
\and
D.~de~Martino\inst{2}
\and
J.-M.~Bonnet-Bidaud\inst{3}
\and
M.~Mouchet\inst{4}
\and
B.~T.~G\"{a}nsicke\inst{5}
\and
G.~Matt\inst{6}
\and
K.~Mukai\inst{7}}


\institute{
Dipartimento di Astronomia, Universit\`{a} di Padova, vicolo dell'Osservatorio 3,
I-35122 Padova, Italy\\
\email{gabriele.anzolin@unipd.it}
\and
INAF - Osservatorio Astronomico di Capodimonte, salita Moiariello 16, I-80131 Napoli, Italy \\
\email{demartino@oacn.inaf.it}
\and
Service d'Astrophysique, DSM/DAPNIA/SAp CE Saclay, F-91191 Gif-sur-Yvette, France \\
\email{bonnetbidaud@cea.fr}
\and
Laboratoire Astroparticule et Cosmologie, Universit\'{e} Paris 7, 10 rue Alice Domon et
L\'{e}onie Duquet, F-75005 Paris, France and LUTH, Observatoire de Paris, Section de Meudon,
5 place Jules Janssen, F-92195 Meudon, France  \\
\email{martine.mouchet@obspm.fr}
\and
Department of Physics, University of Warwick, Coventry CV4 7AL, UK \\
\email{boris.gaensicke@warwick.ac.uk}
\and
Dipartimento di Fisica, Universit\`{a} degli Studi Roma Tre, via della Vasca Navale 84,
I-00146 Roma, Italy \\
\email{matt@fis.uniroma3.it}
\and
CRESST and X-Ray Astrophysics Laboratory, NASA Goddard Space Flight Center, Greenbelt, MD 20771,
USA and Department of Physics, University of Maryland, Baltimore County, 1000 Hilltop Circle,
Baltimore, MD 21250, USA \\
\email{koji.mukai@nasa.gov}
}

\date{Received ...; accepted ...}

\abstract
{}
{We analyze the  first X-ray observations with XMM-Newton of 1RXS~J070407.9+262501 and
1RXS~180340.0+401214, in order to characterize their broad-band temporal and spectral properties,
also in the UV/optical domain, and to confirm them as Intermediate Polars.}
{For both objects, we performed a timing analysis of the X-ray and UV/optical light curves
to detect the white dwarf spin pulsations and study their energy dependence. For 1RXS~180340.0+401214
we also analyzed optical spectroscopic data to determine the orbital period. X-ray spectra were
analyzed in the 0.2--10.0~keV range to characterize the emission properties of both sources.}
{We find that the X-ray light curves of both systems are energy dependent and are dominated,
below 3--5~keV, by strong pulsations at the white dwarf rotational periods (480~s for
1RXS~J070407.9+262501 and 1520.5~s for 1RXS~180340.0+401214). In 1RXS~180340.0+401214 we also detect
an X-ray beat variability at 1697~s which, together with our new optical spectroscopy, favours 
an orbital period of 4.4~hr that is longer than previously estimated.
Both systems show complex spectra with a hard (up to 40~keV) optically thin and a soft
(85--100~eV) optically thick components heavily absorbed by material partially covering the X-ray
sources.}
{Our observations confirm the two systems as Intermediate Polars and also add them as new members
of the growing group of 'soft' systems which show the presence of a soft X-ray blackbody component.
Differences in the temperatures of the blackbodies are qualitatively explained in terms of
reprocessing over different sizes of the white dwarf spot. We suggest that systems showing cooler
soft X-ray blackbody components also possess white dwarfs irradiated by cyclotron radiation.}


\maketitle

\section{Introduction}

Intermediate polars (IPs) are a subclass of magnetic cataclysmic variables (CVs) in which the
rotational period $P_{\omega}$ of the white dwarf (WD) is not synchronized with the orbital period
$P_{\Omega}$ of the binary system ($P_{\omega} \ll P_{\Omega}$). The detection of a modulation
at the WD spin period in the X-ray regime represents a signature of magnetic accretion. The
material lost from the secondary star generally passes through a disc and is channelled onto
the magnetic polar regions of the WD where a strong shock develops \citep{aizu73}. Hard X-ray
radiation is emitted by the hot post-shock gas that cools via thermal bremsstrahlung while it
settles onto the WD surface. Because of the highly asynchronous WD rotation and of the lack of
detectable polarized emission in the optical and near-IR in most systems, IPs were believed
to possess weakly magnetized WDs ($B \sim 0.1-10\ \mathrm{MG}$, \citet{warner}). A distinct
soft ($k T \sim 20-50\ \mathrm{eV}$) X-ray blackbody emission was discovered by the \emph{ROSAT}
satellite in a handful of IPs \citep{mason92,haberletal94,haberlmotch95,burwitzetal96}, some of
them also showing polarized optical radiation indicating field strengths up to 20--30~MG
(e.~g. \citet{piirolaetal93,piirolaetal08}). This spectral
component resembles that observed in the X-ray spectra of Polars, the other subclass of magnetic
CVs, that instead possess synchronously rotating and  strongly magnetized ($B \sim
10 - 230\ \mathrm{MG}$) WDs.

Our current understanding of the origin of soft X-ray emission and energy balance in Polars has
greatly improved since the early model of \citet{lamb_masters79}. It has become evident that the
soft and hard X-ray emissions in at least these systems are largely decoupled
\citep[see][]{beuermann99,beuermann04}, most of the soft X-rays arising from accretion of dense
blobs (`blobby accretion') which penetrate deep in the WD atmosphere. Irradiation of the WD
atmospheric polar regions by hard X-rays and cyclotron radiation emitted in a stand-off shock
has been recently discussed by \citet{konig06}, yelding to the conclusion that most of the
reprocessed light appears in the far-UV rather than in the soft X-rays. 

The similarity with the Polars raised the question on whether the `soft IPs' are  the `true'
progenitors of low-field Polars (see also \citet{norton04}). However, very recently a soft and
rather absorbed X-ray emission is being found in a growing number of IPs
\citep{haberl02,demartino04,demartino06a,demartino06b,demartino08,evanshellier07,staude08}.
Whether all IPs possess such a component and whether it balances the primary emission from the
post-shock flow is a new aspect that deserves investigation. 

Furthermore, in recent years the number of candidate members of the IP group has grown quite
rapidly \citep{woudtwarner04,gan05} almost doubling the number of previously known systems as
well as enlarging the range of asynchronism. While this could suggest an evolution towards
synchronism \citep{norton04}, the confirmation of new candidates is essential to understand the
evolutionary link between IPs and Polars.
 
1RXS~J070407.9+262501 (hereafter RXJ0704) was listed in the ROSAT All-Sky Survey Source Catalogue
(RASS, \citet{vog99}) as a faint X-ray source, with a count rate of $0.19\ \mathrm{cts\ s}^{-1}$
in the 0.1--2.4~keV PSPC range and ROSAT hardness ratios~\footnote{ROSAT hardness ratios 1 and 2
are defined as $\mathrm{HR1} = \frac{(0.40 - 2.40) - (0.07 - 0.40)}{(0.07 - 2.40)}$ and $\mathrm{HR2}
= \frac{(1.00 - 2.40) - (0.40 - 1.00)}{(0.40 - 2.40)}$, where $(A - B)$ is the raw count rate in the
$A - B$ energy range expressed in keV \citep{motchetal96}.} $\mathrm{HR1} = -0.40$ and
$\mathrm{HR2} = +0.55$ that suggested the presence of a soft component in the X-ray spectrum.
RXJ0704 was then identified as a CV by \citet{gan05} and proposed to be an IP from the
detection of 
optical photometric and spectroscopic periodicities at 480.71~s and $\sim 250\ \mathrm{min}$,
respectively. The short period variability was ascribed to the spin period of the WD, while the
other was identified as the orbital period of the binary system. The optical light curve folded at
the putative WD rotational period was double peaked, suggesting the presence of two equally
contributing accreting poles. However, no other X-ray observation was performed on this source in
order to study its X-ray properties and hence to confirm its membership in the IP class.

1RXS~180340.0+401214 (hereafter RXJ1803) was also discovered in the RASS as a faint (PSPC count
rate $0.18\ \mathrm{cts\ s}^{-1}$) source with a very soft X-ray spectrum ($\mathrm{HR1} =
-0.18$ and $\mathrm{HR2} = -0.02$). Later, it was also identified as an IP candidate by \citet{gan05}
from its optical photometric variability at a 1520.5~s period that was interpreted as the rotational
period of the accreting WD. A spectroscopic period of 160.2~min was inferred from the analysis
of radial velocity variations in the $\mathrm{H}_{\alpha}$ emission line wings and identified as
the binary orbital period, thus placing this CV in the 2--3~hr period gap. The two periods would
then indicate a system with a weak degree of asynchronism. Also for this source no further X-ray
observations were available.

In this work we report XMM-Newton observations of RXJ0704 and RX1803 aiming at determining
for the first time their X-ray variability as well as their X-ray broad-band spectral properties.
For RXJ1803 we further complement the XMM-Newton data with optical spectroscopy acquired at the
Observatoire de Haute Provence (OHP, France).

\section{The observations and data reduction}

\begin{table*}
\caption{Summary of the XMM-Newton observations of RXJ0704 and RXJ1803}
\label{tab:observ}
\centering 
\begin{tabular}{llllll}
\hline\hline 
Object	& Instrument	& Date			& UT (start)	& Net exposure time (s)	&
Net count rate ($\mathrm{cts\ s}^{-1}$)\\
\hline
RXJ0704	& EPIC-pn		& 2006-10-04    & 05:52			& 10194		& $1.036 \pm 0.010$ \\
		& EPIC-MOS		&				& 05:22			& 12240		& $0.232 \pm 0.005$ \\
		& RGS			&				& 04:59 		& 13908		& $0.032 \pm 0.002$ \\
		& EPIC-pn		& 2007-03-23	& 16:03			& 2900		& $1.463 \pm 0.023$	\\
		& EPIC-MOS		&				& 15:41			& 4450		& $0.364 \pm 0.011$	\\
		& RGS			&				& 15:40			& 4721		& $0.047 \pm 0.001$ \\
		& OM-UVM2		&				& 15:49			& 1239		& $0.749 \pm 0.025$	\\
		&				&				& 16:45			& 1239		& $0.822 \pm 0.026$	\\
		&				&				& 17:11			& 1240		& $0.669 \pm 0.023$	\\
		&				&				& 17:37			& 1241		& $0.634 \pm 0.023$	\\
		&				&				& 18:03			& 1241		& $0.630 \pm 0.023$	\\
\hline
RXJ1803	& EPIC-pn 		& 2007-08-31 	& 02:18			& 18457		& $0.503 \pm 0.005$ \\ 
        & EPIC-MOS		&      			& 01:56			& 19356		& $0.148 \pm 0.003$ \\ 
        & RGS			&    			& 01:56			& 18737		& $0.020 \pm 0.001$ \\ 
		& OM-B    		&    			& 02:05			& 1839		& $1.645 \pm 0.030$ \\
		&				&				& 02:41			& 1840		& $1.808 \pm 0.031$	\\
		&				&				& 03:17			& 1842		& $1.850 \pm 0.032$ \\
		&				&				& 03:53			& 1841		& $2.040 \pm 0.033$ \\
		&				&				& 04:29			& 1839		& $2.076 \pm 0.034$ \\
		& OM-UVM2		&				& 05:05			& 1840		& $0.263 \pm 0.012$ \\
		&				&				& 05:41			& 1840		& $0.239 \pm 0.011$ \\
		&				&				& 06:17			& 1839		& $0.222 \pm 0.011$ \\
		&				&				& 07:29			& 1839		& $0.255 \pm 0.012$ \\
\hline\end{tabular}
\end{table*}

\subsection{The XMM-Newton observations of RXJ0704}

RXJ0704 was observed with the XMM-Newton observatory \citep{jan01} on October 4, 2006
(OBSID: 0401650101) with the EPIC-pn \citep{str01} and MOS \citep{tur01} cameras operating in
full frame mode with the medium filter for total exposure times of 10.2~ks and 12.2~ks,
respectively. The RGS instruments \citep{den01} were operated in spectroscopy mode for an exposure
time of 13.9~ks. No useful data were collected with the OM \citep{mas01}. Due to a failure of
the XMM-Newton mission operation centre system, the exposure was interrupted. RXJ0704 was then
observed again on March 23, 2007 (OBSID: 0401650301) with the same instrument configurations.
The OM was operated in fast imaging mode using the UVM2 filter that covers the ultraviolet spectral
range 2000--2800~\AA. The total exposures were 8.1~ks for EPIC-pn, 9.6~ks for the MOS, 9.9~ks for
the RGS and 6.2~ks for the OM. A summary of the observations is reported in Table~\ref{tab:observ}.
All the raw data have been reprocessed using the standard pipeline process included in the
SAS~7.0 package. EPIC light curves and spectra were extracted using a circular region of
30\arcsec centred on the source. The same aperture radius was used to extract background light
curves and spectra on the same CCD chip of the target. To improve S/N, we selected up to double
pixel events with zero quality flag for the EPIC-pn data, while for the EPIC-MOS cameras we
increased the selection level up to quadruple pixel events.

The EPIC data of October 2006 were characterized by a very low and rather constant background
level ($0.014\ \mathrm{cts\ s}^{-1}$ and $0.002\ \mathrm{cts\ s}^{-1}$ for EPIC-pn and MOS
respectively). In order to obtain sufficient S/N ratios and use $\chi^2$ statistics in the
spectral analysis, the average EPIC spectra were rebinned to obtain 20 counts in each spectral bin.
The photon redistribution matrix and the ancillary region file were created by using the SAS tasks
\emph{rmfgen} and \emph{arfgen}. The RGS1 and RGS2 first order spectra are too faint to allow
a reliable analysis and, therefore, are not presented in this work. 

The March 2007 X-ray data were instead severely plagued by high background radiation for more
than half of the duration of the observation. We then windowed the EPIC data to exclude those
epochs of high ($> 0.5\ \mathrm{cts\ s}^{-1}$) background count rate, leaving us with the initial
2.9~ks of useful data, thus losing $\sim 70\%$ of the whole observation. The extracted EPIC
light curves and average spectra were obtained using the same procedure as described before. Also
in this case the RGS data did not provide an useful spectrum. The OM-UVM2 background subtracted 
light curve binned in 20~s intervals was obtained with the task \emph{omfchain}. The average 
count rate was $0.7\ \mathrm{cts\ s}^{-1}$, corresponding to an instrumental magnitude
$\mathrm{UVM2} = 16.2$ and to an average flux of
$1.5 \times 10^{-15}\ \mathrm{erg\ cm}^{-2} \mathrm{s}^{-1}\ \mbox{\AA}^{-1}$ in the
2000--2800~\AA\ band. For a comparison, the continuum flux of the optical spectrum obtained
by \citet{gan05} is $\sim 2.2 \times 10^{-15}\ \mathrm{erg\ cm}^{-2}\ \mathrm{s}^{-1}\ \mbox{\AA}^{-1}$
at 3800~\AA.

\subsection{The XMM-Newton observations of RXJ1803}

RXJ1803 was observed on August 31, 2007 (OBSID:0501230101) with all EPIC cameras operated in full
frame mode and with the thin filter. The total exposure times were 20.0~ks and 21.7~ks for the EPIC-pn
and MOS cameras, respectively. Of the two RGS instruments, only RGS1 could acquire data for 21.9~ks,
while RGS2 was blocked in a `set-up' mode during the observation. The OM was operated in fast imaging
mode using sequentially the UVM2 and B filters, the latter covering the spectral range 3900--4900~\AA,
for 9.2~ks each (see Table~\ref{tab:observ}). The EPIC data were reprocessed and, then, filtered to
avoid a flare event that occurred during the initial $\sim 2000\ \mathrm{s}$ of the observation. Light curves
and spectra were extracted using a radius of 30\arcsec centered on the source. We also applied similar
pixel selection levels as done before for the EPIC-pn and MOS cameras and rebinned the spectra to
obtain 25 counts per bin. Also for this source, the RGS1 first order spectrum was too faint to
allow useful analysis. Due to a failure in the \emph{omfchain} routine of SAS~7.0, OM data were
reprocessed at the MSSL using new, yet unreleased, routines. Unfortunately, there was still
a processing problem in one of the exposures with the UVM2 filter that cannot be solved. However,
we obtained OM-B and OM-UVM2 background subtracted light curve binned in 10~s intervals. The average
count rates were $1.88\ \mathrm{cts\ s}^{-1}$ in the B band and $0.24\ \mathrm{cts\ s}^{-1}$ in the
UVM2 band. The corresponding instrumental magnitude were $\mathrm{B} = 18.6$ and $\mathrm{UVM2} = 17.3$,
while the average fluxes were
$2.4 \times 10^{-16}\ \mathrm{erg\ cm}^{-2}\ \mathrm{s}^{-1}\ \mbox{\AA}^{-1}$
and $5.2 \times 10^{-16}\ \mathrm{erg\ cm}^{-2}\ \mathrm{s}^{-1}\ \mbox{\AA}^{-1}$, respectively.

Heliocentric corrections were applied to the EPIC and OM data of both sources.

\subsection{OHP spectroscopy of RXJ1803}

RXJ1803 was observed at the OHP during two nights on June 29 and July 1, 2006 for 4.3~hr and
4.22~hr, respectively.
Long slit spectra were obtained with the Carelec spectrograph \citep{lemaitre90} placed at
the Cassegrain focus of the 193~cm telescope and using a CCD EEV ($2048 \times 1024$ pixels) detector
with $13.5\ \mu\mathrm{m}$ pixel size. A $133\ \mbox{\AA} / \mathrm{mm}$ grating was used, with a slit
width of 2\arcsec, leading to a wavelength coverage of 3600--7200~\AA\ at a FWHM resolution of
$\sim 5.7\ \mbox{\AA}$. All exposure times were set at 1520~s, corresponding to the known optical
pulsation, to smear any effect from the spin variability. The observations were performed
in nearly photometric conditions with seeing typically ranging from $2 \farcs 5$ to 3\arcsec.
We obtained 9 spectra of RXJ1803 each night. Standard reduction
was performed with the ESO-MIDAS package, including cosmic rays removal, bias subtraction, flat-field
correction and wavelength calibration. The wavelength calibration was checked on sky lines and all
radial velocity measurements have been corrected from the small instrumental shifts measured on
the \ion{O}{I} 5577~\AA\ line and from the Earth motion. Times have been converted in the
heliocentric system. Flux calibration has been performed using the standard stars BD+28~4211
and BD~+33 2642 .

\section{Analysis and results}

\subsection{RXJ0704}

We have analysed the X-ray and UV XMM-Newton data of RXJ0704 to search for variability at the
optical period as well as to study its broad band spectral properties.

\begin{figure}
\centering
\resizebox{\hsize}{!}{\includegraphics{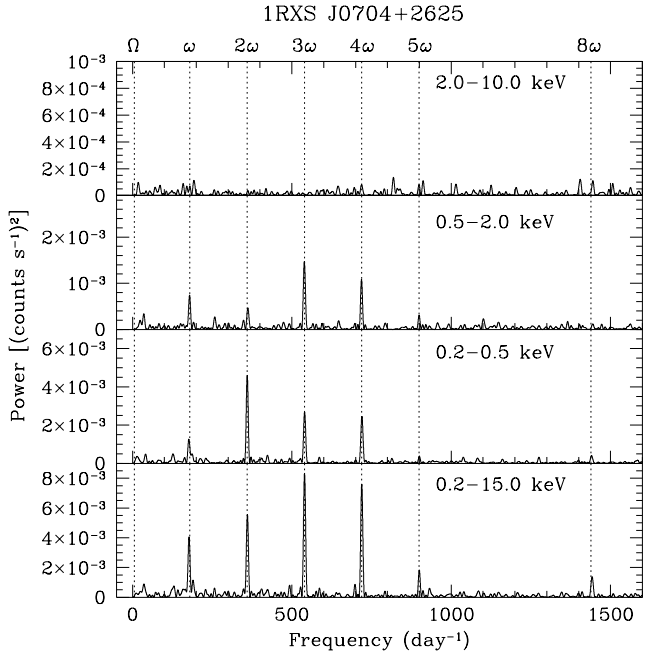}}
\caption{Power spectra of the October 2006 EPIC-pn data of RXJ0704 in selected energy ranges.
\emph{From bottom to top}: 0.2--15.0~keV, 0.2--0.5~keV,  0.5--2.0~keV, 2.0--10.0~keV. The spin
($\omega$) frequency, its harmonics and the orbital frequency ($\Omega$) are marked with vertical
dotted lines.}
\label{fig:powoct}
\end{figure}

\begin{figure}
\centering
\resizebox{\hsize}{!}{\includegraphics{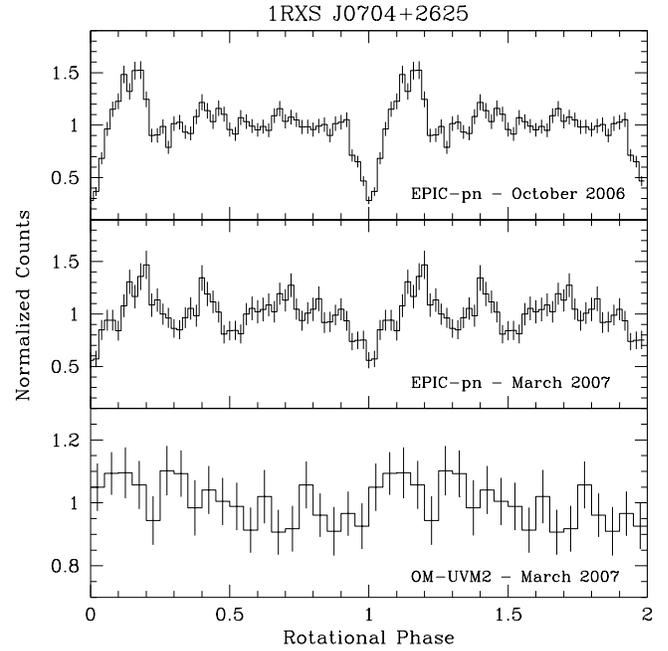}}
\caption{Light curves of RXJ0704 folded at the 480~s spin period using the times of the dip quoted in
the text. \emph{Top and central panels}: full-band (0.2--15.0~keV) EPIC-pn light curves for October
2006 and March 2007. \emph{Bottom panel}: the OM-UVM2 spin pulsation of March 2007 detrended from
the low frequency variability (see text for details).}
\label{fig:flc-all}
\end{figure}

\subsubsection{The X-ray variability}

The light curves from all EPIC cameras were extracted in the range 0.2--15.0~keV with a 20~s time
resolution. A clear signature of a periodicity on a time scale of a few dozen of minutes is detected
in both the October 2006 and March 2007 EPIC-pn data. Since EPIC-MOS light curves are very similar
to those obtained with the EPIC-pn camera in both epochs, though with lower statistics, we only show
the EPIC-pn data analysis. Fourier analysis was restricted only to the October 2006 data
(Fig.~\ref{fig:powoct}) because of the short duration of the March 2007 observations. We found
significant peaks at $\omega \sim 180\ \mathrm{day}^{-1}$ and at $2 \omega$, but the strongest peaks
occur at $3 \omega$ and $4 \omega$. There are also weak signals at $5 \omega$ and $8 \omega$,
indicating a very structured X-ray light curve. A fit was performed on the October 2006 EPIC-pn
and MOS light curves by using a composite function with up to six sinusoids to account for all
the detected harmonics. We let the fundamental frequency $\omega$ free to vary, while the other five
frequencies were constrained to be integer multiple values of $\omega$, i.~e. $2\omega$, $3\omega$,
$4\omega$, $5\omega$ and $8\omega$. In this way, we found $\omega = 179.88 \pm 0.06\ \mathrm{day}^{-1}$
for EPIC-pn, $179.7 \pm 0.3\ \mathrm{day}^{-1}$ for MOS1 and $179.85 \pm 0.15\ \mathrm{day}^{-1}$ for
MOS2. Because of the higher accuracy of the EPIC-pn data, we adopt $P_{\omega} = 480.3 \pm 0.2\ \mathrm{s}$.
This value is consistent within $3 \sigma$ with the optical photometric determination of \citet{gan05}
and hence can be safely ascribed to the WD rotational period. 

Fig.~\ref{fig:flc-all} shows the full band EPIC-pn light curves of both epochs folded at the
spin period. The light curves appear very complex and are dominated by two features: a dip, that
we arbitrarily placed at phase zero, and a bump at phase $\sim 0.15$. The dip represents a
fiducial marker of the spin pulsation. We hence derive the following times of the dip:
$\mathrm{HJD} = 2454012.74140(7)$ for October 2006 and $\mathrm{HJD} = 2454183.17362(5)$ for March
2007. The spin pulse is more structured in March 2007, especially in the phase range $0.15-0.5$.
The dip and bump appear to be less defined, while an additional peak can be recognized at
$\phi \sim 0.4$. This latter could also be present in the October 2006 data, but with much
lower amplitude.

\begin{figure}
\centering
\resizebox{\hsize}{!}{\includegraphics{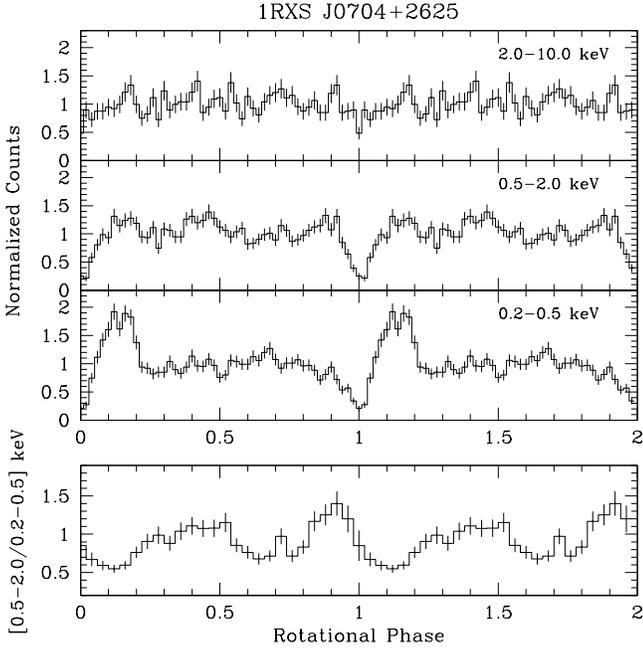}}
\caption{RXJ0704 data of October 2006. \emph{Top panels}: EPIC-pn light curves in selected
energy ranges folded at the 480~s spin period using the time of the dip quoted in the
text. \textit{Bottom panel}: EPIC-pn hardness ratio in the soft bands.}
\label{fig:flc-hrt_oct}
\end{figure}

The Fourier analysis was also performed on the October 2006 EPIC-pn light curves extracted with
the same 20~s binning time in selected energy bands: 0.2--0.5~keV, 0.5--2.0~keV and 2.0--10.0~keV
(see Fig.~\ref{fig:powoct}). While no sign of variability can be detected  above 2.0~keV,
significant peaks are evident in the soft band periodograms, where the spin frequency and its
first three harmonics are clearly detected. The October 2006 folded light curves are shown in
Fig.~\ref{fig:flc-hrt_oct}. The dip feature is clearly observed below 2~keV but the bump appears
only below 0.5~keV. The March 2007 energy dependent light curves confirm the general behaviour
observed in October 2006, with no sign of variability above 2~keV and a dip whose depth increases
at lower energies. The fractional depth (relative to the average count rate level) of the dip is,
within errors, slightly larger in October 2006 than in March 2007 ($75 \pm 7\%$ and $66 \pm 12\%$,
respectively). Similarly the bump centered at phase $\sim 0.15$ has a fractional amplitude of
$87 \pm 14\%$ in October 2006, while its amplitude is $62 \pm 10\%$ in March 2007. Both the
dip and the bump are extremely localized in phase ($\Delta \phi \sim 0.1$), therefore the regions
responsible for these features should have a very small extension.

In the lower panel of Fig.~\ref{fig:flc-hrt_oct} we also plot the hardness ratio defined as 
the ratio between the 0.5--2.0~keV and 0.2--0.5~keV count rates (the hardness ratio obtained 
with the other energy bands does not provide useful information). The count rate ratio of the 
soft bands is double-humped showing a hardening during the dip as well as at $\phi \sim 
0.3-0.5$ that corresponds to the `flat' part of the spin pulse. This clearly reflects the presence
of the soft bump and the dip.

\subsubsection{The UV light curve} \label{sec:uv0704} 

The March 2007 OM-UVM2 light curve shows hints of a quasi-sinusoidal modulation on the order
of few hours. However, this cannot be quantified because of the short duration
of the observation. To better study the short term variability we removed the long term variations
from the UV light curve using a third order polynomial and folded the
residual at the X-ray spin period with the epoch of the dip found for March 2007. As shown in the lower
panel of Fig.~\ref{fig:flc-all}, the pulsation is quasi-sinusoidal. A sinusoidal fit gives the maximum
at phase $0.22 \pm 0.03$, hence partially overlapping the X-ray bump and the additional peak at
$\phi \sim 0.4$ observed at that epoch, and a modulation fraction of $12 \pm 5\%$ (defined as $2 A
/ f_{\mathrm{ave}}$, where $A$ is the amplitude of the fitted sinusoid and $f_{\mathrm{ave}}$ is the
average count rate). We also recall that the white light optical pulsation observed by \citet{gan05}
was instead found to be double-humped.

\subsubsection{Spectral properties of RXJ0704}

\begin{figure}
\centering
\resizebox{\hsize}{!}{\includegraphics{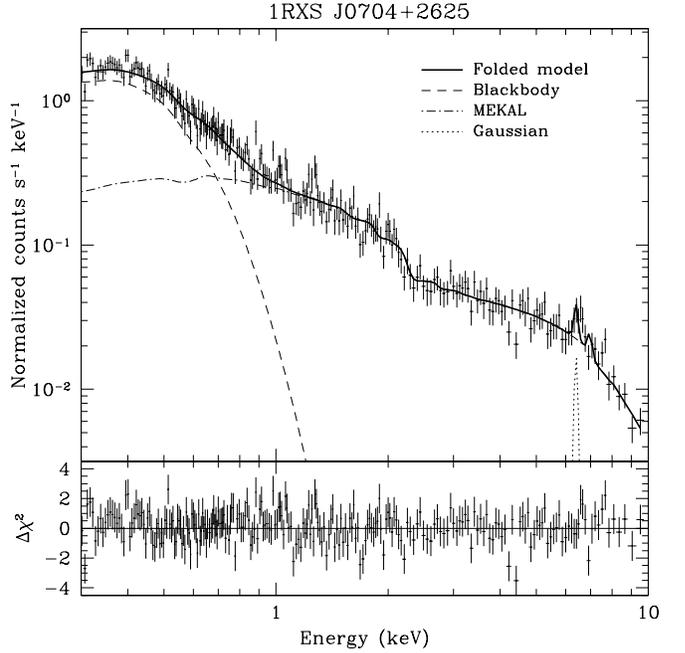}}
\caption{The October 2006 EPIC-pn spectrum of RXJ0704 is shown with the best fit model discussed
in the text and parameters reported in Table~\ref{tab:spectra0704} (for the sake of simplicity,
we do not plot the combined MOS spectrum). Each spectral component that contributes to the best-fit
model is shown separately, for clarity. The bottom panel shows the residuals expressed in terms
of $\sigma$.}
\label{fig:specoct}
\end{figure}

The October 2006 (Fig.~\ref{fig:specoct}) and March 2007 average EPIC-pn and combined MOS spectra
were analyzed with the XSPEC~12 package in the range 0.3--10.0~keV. A simple model consisting of
an absorbed optically thin plasma MEKAL with solar abundances does not adequately fit the spectra.
For October 2006 data, a zero-width Gaussian centered at 6.4~keV is also required, although the
fit is still poor ($\chi_{\nu}^2 = 8.8$). Both October 2006 and March 2007 spectra
show an excess of counts below 1~keV and the addition of a blackbody ($k T_{\mathrm{BB}} \sim
80\ \mathrm{eV}$) drastically improves the fits ($\chi_{\nu}^2  = 0.97$ and 0.70, respectively).
For the October 2006 spectrum, the temperature of the optically thin plasma is unconstrained
($k T_{\mathrm{MEK}} > 44\ \mathrm{keV}$), while for the March 2007 spectrum it is $k T_{\mathrm{MEK}}
= 11 \mathrm{keV}$. However, a multi-temperature plasma, expected to be present in the post-shock
region, does not improve the fit quality. If the abundance is left free to vary, it assumes an
unrealistic value $A_{\mathrm{Z}} = 1.7$ for the October 2006 data, while for the March 2007
spectrum we obtain $A_{\mathrm{Z}} = 0.6$, but still within errors compatible with solar abundances.
We therefore fix $A_{\mathrm{Z}}$ to the solar value. The fits further improve
by including an additional
dense absorber covering the source by $\sim 40\%$, as detailed in Table~\ref{tab:spectra0704}. The
inclusion of a denser absorber is justified by a F-test at significance level of 99.9\%. The low column
density of the total absorber, when compared with the total galactic absorption in the direction of
the source ($N_{\mathrm{H,\,gal}} = 8.15 \times 10^{20}\ \mathrm{cm}^{-2}$, \citet{dic90}), suggests a
likely interstellar origin. On the other hand, the dense absorber should be confined within the
binary system and likely due to pre-shock material as suggested by the energy dependence of the
spin light curve. No major differences are found in the spectral parameters between the October
2006 and March 2007 data (see Table~\ref{tab:spectra0704}), except for the lack of detectable iron
6.4~keV line in the March 2007 low S/N spectrum and a significant change in the normalization of the
blackbody component, which increases by a factor of two in March 2007. This change could explain the 
increase in the source  flux observed between the two epochs
($F_{0.2-10.0\ \mathrm{keV}}^{\mathrm{Oct.}} = 4.2 \times 10^{-12}\ \mathrm{erg\ cm}^{-2}\ \mathrm{s}^{-1}$
and
$F_{0.2-10.0\ \mathrm{keV}}^{\mathrm{Mar.}} = 5.3 \times 10^{-12}\ \mathrm{erg\ cm}^{-2}\ \mathrm{ s}^{-1}$,
respectively).

\begin{table*}
\begin{minipage}[t]{\hsize}
\caption{Spectral parameters of the best-fit model to the EPIC-pn and combined MOS average spectra
of RXJ0704. Errors indicate the 90\% confidence level of the relevant parameter. Metal abundance
is kept fixed at the solar value (see text).}
\label{tab:spectra0704} 
\centering
\renewcommand{\footnoterule}{}
\begin{tabular}{llllllllll}
\hline\hline       
Epoch & Total Absorber & \multicolumn{2}{c}{Partial Absorber} & \multicolumn{2}{c}{Blackbody} &
 \multicolumn{2}{c}{MEKAL} & Iron Line & $\chi_{\nu}^2$ ($\chi^2$ / d.o.f.) \\
& $N_{\mathrm{H}}$~\footnote{Column density of the total absorber.} &
 $N_{\mathrm{H}}$~\footnote{Column density of the partial absorber.} & Cov.
 Frac.~\footnote{Covering fraction of the partial absorber.} & $k T_{\mathrm{BB}}$ &
 $N_{\mathrm{BB}}$~\footnote{Normalization constant of the BBODY model.} & $k T_{\mathrm{MEK}}$ &
 $N_{\mathrm{MEK}}$~\footnote{Normalization constant of the MEKAL model.} &
 EW~\footnote{Equivalent width of the 6.4 keV iron line.} & \\
& $(10^{20}\ \mathrm{cm}^{-2})$ & $(10^{23}\ \mathrm{cm}^{-2})$  & & (eV) & $(10^{-5})$ & (keV) &
 $(10^{-3})$ & (eV) \\
\hline
Oct. 2006 & $1.0_{-0.5}^{+0.6}$ & $1.9_{-0.9}^{+1.9}$ & $0.33_{-0.08}^{+0.13} $ & $84 \pm 3$ &
 $3.0_{-0.4}^{+0.9}$ & $> 44$ & $2.3 \pm 0.3$ & $141 \pm 73$ & 0.87 (360/415) 
\\
Mar. 2007 & $0.5 \pm 0.4$ & $1.2_{-0.6}^{+1.2}$ & $0.54_{-0.09}^{+0.07}$ & $88 \pm 5$ &
 $6.3_{-1.6}^{+1.7}$ & $11_{-3}^{+13}$ &$2.6_{-0.5}^{+0.4}$ &  & 0.63 (121/193) \\
\hline
\end{tabular}
\end{minipage}
\end{table*}

We also performed a phase-resolved analysis of the October 2006 EPIC-pn spectra (the inclusion of
the MOS spectra does not greatly improve the statistics) extracted around the very localized dip
($\phi = 0.9-1.05$) and bump ($\phi = 0.1-0.2$), as well as in the flat part ($\phi = 0.3-0.8$). 
Since the X-rays are essentially unpulsed above 2\,keV, we used the best-fit model of October 2006
keeping fixed the MEKAL parameters to the values found for the average spectrum. We also fixed the
column density of total absorber as this is not expected to vary along the spin cycle. The Gaussian
component used to account for the iron line at 6.4~keV is not required to fit the spectra extracted
arount the dip and the bump because of the poor S/N ratio. Unfortunately, with our data many spectral
parameters are badly constrained, as shown in Table~\ref{tab:phspectra}. Anyhow, we find a
substantial change in the covering fraction of the partial absorber between the flat portion of the
pulse and both the dip and the bump. The values of the normalization of the blackbody component found
in the flat part and in the dip are consistent, within the errors, but there is an evident increase
in the bump. Hence, it is conceivable that the spin pulse of RXJ0704 is due to an increase in the local
absorption (dip feature) and to an increase in the projected area of the blackbody component (bump).

\begin{table}
\caption{Spectral parameters obtained from fitting the October 2006 phase-resolved EPIC-pn spectra
of RXJ0704 extracted in the phase intervals quoted in the text with the model shown in
Table~\ref{tab:spectra0704}.} 
\label{tab:phspectra}
\centering
\begin{tabular}{llll}
\hline\hline
Parameters 					  	& Dip				& Bump						& Flat \\
\hline
$F_{0.2-10.0\ \mathrm{keV}}$	& 2.71 				& 5.15 						& 4.13 \\
($10^{-12}\ \mathrm{erg\ cm}^{-2}\ \mathrm{ s}^{-1}$) &	&						& \\
\hline
$N_{\mathrm{H}}$ ($10^{23}\ \mathrm{cm}^{-2}$) & $2.1_{-0.8}^{+1.3}$ & $1.3_{-0.7}^{+1.8}$ & $2.7_{-0.7}^{+1.1}$ \\
Cov. Frac.						& $0.73 \pm 0.04$	& $0.48_{-0.07}^{+0.06}$	& $0.44 \pm 0.20$ \\
$k T_{\mathrm{BB}}$ (eV)		& $84_{-10}^{+11}$	& $82 \pm 4$				& $85 \pm 2$ \\
$N_{\mathrm{BB}}$ ($10^{-5}$)	& $3.3_{-0.8}^{+1.0}$ & $7.7_{-1.0}^{+1.3}$		& $3.6 \pm 0.2$ \\
\hline
$\chi_{\nu}^2$ ($\chi^2$ / d.o.f.) & 1.15 (26/23)	& 1.03 (63/61)				& 1.00 (297/298) \\
\hline
\end{tabular}
\end{table}

\subsection{RXJ1803}

Similarly to RXJ0704, the XMM-Newton data were also analyzed to search for periodicities in
the X-ray, UV and optical ranges. The spectroscopy acquired at OHP was analyzed to search
for variability in the radial velocities of Balmer lines.

\begin{figure}
\centering
\resizebox{\hsize}{!}{\includegraphics{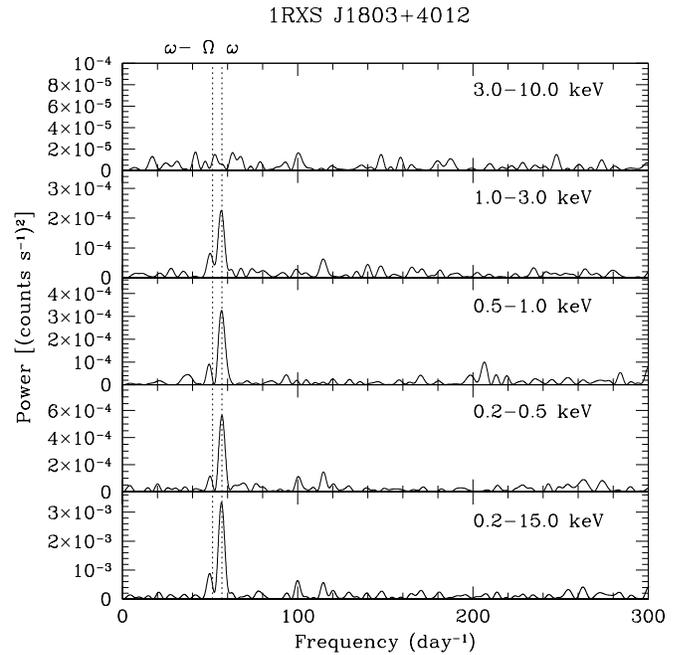}}
 \caption{Power spectra of the EPIC-pn light curve of RXJ1803 in selected energy ranges.
\emph{From bottom to top}: 0.2--15.0~keV, 0.2--0.5~keV, 0.5--1.0~keV, 1.0--3.0~keV and
3.0-10.0~keV. The spin ($\omega$), and the orbital sideband ($\omega - \Omega$) as determined
by optical photometry and our spectroscopy (see text) are marked with vertical dotted lines.}
\label{fig:rxj1803_powspec}
\end{figure}

\subsubsection{The X-ray periodicities in RXJ1803}

The EPIC-pn light curve (EPIC-MOS light curves were too noisy to provide useful information) was
extracted in the full energy range of the camera with a resolution of 20~s and Fourier analyzed.
The power spectra in the 0.2--15.0~keV range, reported in Fig.~\ref{fig:rxj1803_powspec}, show the
presence of a strong peak at $56.6\ \mathrm{day}^{-1}$ and an additional weaker peak at
$49.8\ \mathrm{day}^{-1}$. A sinusoidal fit composed with one single frequency gives then a period
of $1528 \pm 6\ \mathrm{s}$. This is fairly consistent with the highly accurate optical period of
$1520.510 \pm 0.066\ \mathrm{s}$ determined by \citet{gan05}, allowing us to confirm it as the
true spin period of the WD. The residual light curve when fitted with a second sinusoid gives a
period of $1697 \pm 22\ \mathrm{s}$. If this frequency represents the sideband $\omega - \Omega$
(commonly named as the beat), using either the X-ray period or the optical period we derive an orbital
frequency of $5.64 \pm 0.69\ \mathrm{day}^{-1}$ or $5.91 \pm 0.66\ \mathrm{day}^{-1}$, respectively.
This gives an orbital period of $4.25 \pm 0.52\ \mathrm{hr}$ or $4.06 \pm 0.45\ \mathrm{hr}$ which
does not match either the orbital period of 2.67~hr determined by \citet{gan05}, or $2 P_{\Omega}$.
As it will be shown below, the orbital period is likely 4.4~hr, hence longer than previously determined.
The detection of X-ray variability at the spin and beat frequencies (with different amplitudes)
is also observed in other IPs \citep[see][]{norton92b,norton92,norton97}. In RXJ1803, the ratio
of power between spin and beat frequencies is $\sim 3.8$, implying that the beat modulation has
an amplitude about twice lower than that of the spin.

\begin{figure}
\centering
\resizebox{\hsize}{!}{\includegraphics{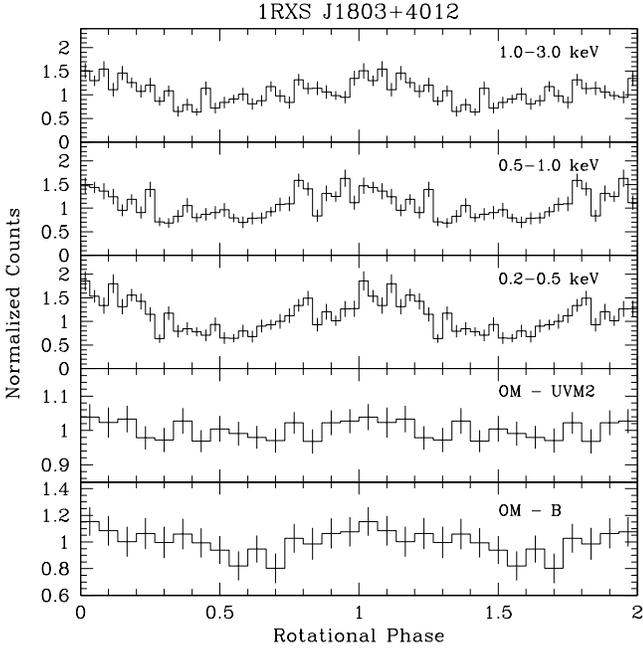}}
\caption{The spin pulse of RXJ1803 in selected energy ranges of the EPIC-pn and in the B and
UVM2 bands. All the light curves are folded at the 1520.5~s optical spin period using the
time of maximum quoted in the text.} 
\label{fig:rxj1803_flc}
\end{figure}

\begin{figure}
\centering
\resizebox{\hsize}{!}{\includegraphics{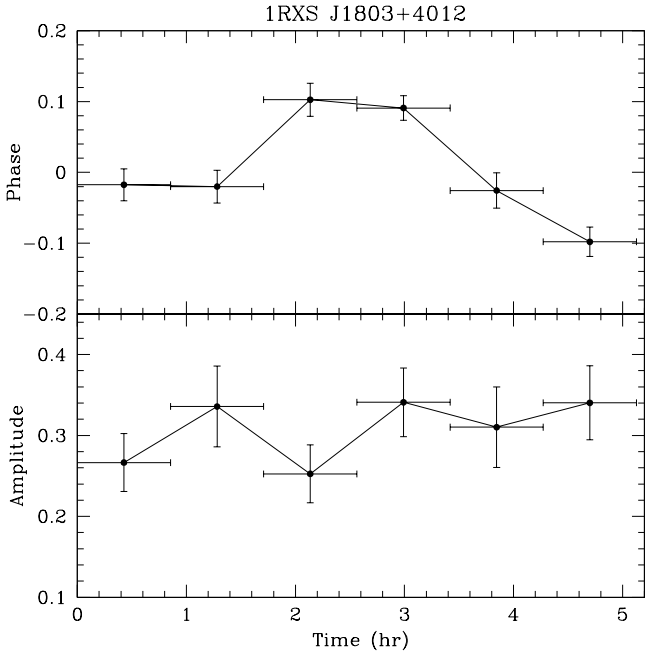}}
\caption{Analysis of the X-ray spin pulse of RXJ1803 in the range 0.2--3.0~keV. \emph{Upper panel}:
the time dependence of the phase of the spin maximum. \emph{Lower panel}: the time dependence of spin
amplitudes.}
\label{fig:rxj1803_beat}
\end{figure}
 
Light curves of the EPIC-pn were also extracted in different energy ranges with a binning time
of 20~s and Fourier analyzed, as also shown in Fig.~\ref{fig:rxj1803_powspec}. The spin variability
is clearly detected below 3~keV, as well as the beat frequency. The ratio of power between spin
and beat peaks is $\sim 4.8$ in the 0.2--0.5~keV, $\sim 3.6$ in the 0.5--1.0~keV and $\sim 2.8$ in
the 1.0--3.0~keV, suggesting that the variabilities at the two frequencies do not have the same
energy dependence. Because of the higher accuracy of the optical period determination, 
we folded the spin light curves at the 1520.5~s period (Fig.~\ref{fig:rxj1803_flc}) using the time
of maximum obtained from a sinusoidal fit with fixed frequency: $\mathrm{HJD} = 2454343.6124(4)$.
The folded light curves are quasi-sinusoidal with a dip at $\phi \sim 0.8$ which is more evident in
the softest 0.2--0.5~keV band. The pulse amplitude (see Sect.~\ref{sec:uv0704}) is $72 \pm 8\%$ in
the 0.2--0.5~keV range, $57 \pm 7\%$ in the 0.5--1.0~keV and $46 \pm 6\%$ in the 1.0-3.0~keV band,
indicating a strong energy dependence especially below 1~keV. The hardness ratios in the three bands
however do not substantially vary along the spin cycle, indicating that the spin pulse is dominated
by aspect angle variations. 

The beat variability suggests that the shape of the spin pulsation is variable along the binary period.
The XMM-Newton observation spans about 5.6~hr, likely covering more than one orbital period. We then
extracted the spin light curve in six intervals each sampling $\sim 3000\ \mathrm{s}$ (i.~e. about two
spin cycles). While amplitudes are rather constant within errors, the spin pulse changes phase with
time (see Fig.~\ref{fig:rxj1803_beat}). Altough we cannot establish whether this variability is periodic,
the presence of the beat variability suggests it might be the case.

\subsubsection{The UV and optical light of RXJ1803}

The OM-UVM2 light curve does not reveal any clear trend, while that in the B filter shows
both a long term and short term variability. Count rates steadily increase by a factor of $\sim 1.4$
during the 2.6~hr of observation with the B filter. This could reflect an orbital dependence,
suggesting, as found in the X-ray data, that the orbital period is longer than previously estimated
\citep{gan05}. We then removed the long term trend in the B-band light curve using a third order
polynomial. The folded spin light curves in the two bands are shown in Fig.~\ref{fig:rxj1803_flc}.
Although very noisy, they appear to be different: in the UV no modulation is apparent within errors,
while the B band light curve is quasi-sinusoidal, with an amplitude of $30 \pm 10\%$. The B band
pulse is single peaked, as also detected by \citet{gan05}, and is broadly consistent in phase with
the X-ray pulse.

\subsubsection{The X-ray spectrum of RXJ1803}

\begin{table*}
\begin{minipage}[t]{\hsize}
\caption{Spectral parameters of the  models used to fit the EPIC-pn and combined MOS average
spectra of RXJ1803, as described in the text. Errors indicate the 90\% confidence level of the
relevant parameter.}
\label{tab:spectra1803} 
\centering
\renewcommand{\footnoterule}{}
\begin{tabular}{lllllllllll}
\hline\hline       
& \multicolumn{2}{c}{Partial Absorber} & \multicolumn{2}{c}{Blackbody} &
 \multicolumn{5}{c}{Optically thin plasma~\footnote{The optically thin plasma is represented by
 a CEMEKL (Models A and B) with fixed slope $\alpha = 1$ or by two MEKALs (Model C).}} &
 $\chi_{\nu}^2$ ($\chi^2$ / d.o.f.) \\
& $N_{\mathrm{H}}$~\footnote{Column density of the partial absorber.} & Cov.
 Frac.~\footnote{Covering fraction of the partial absorber.} & $k T_{\mathrm{BB}}$ &
 $N_{\mathrm{BB}}$~\footnote{Normalization constant of the BBODY model.} &
 $A_\mathrm{Z}$~\footnote{Metal abundance.} & $k T_1$ &  $N_1$~\footnote{Normalization constant of
 the first MEKAL model (only for Model C).} & $k T_2$~\footnote{In Models A and B this is the maximum
 temperature of CEMEKL model.} & $N_2$~\footnote{Normalization constant of the CEMEKL (Models A and B)
 or the second MEKAL model (Model C).} & \\
& $(10^{23}\ \mathrm{cm}^{-2})$ & & (eV) & $(10^{-6})$ & & (keV) & $(10^{-3})$ & (keV) &
 $(10^{-3})$ & \\
\hline
A & & & $98 \pm 8$ & $2.3\pm 0.2$ & $1.2_{-0.3}^{+0.4}$ & & & 40$_{-8}^{+10}$ &
 $1.61_{-0.07}^{+0.09}$ & 0.91 (448/495) \\ 
B & $3_{-1}^{+3}$ & $0.4 \pm 0.2$ & $95_{-9}^{+10}$ & $3.3_{-0.6}^{+1.4}$ & $0.4_{-0.2}^{+0.3}$
 & & & $21_{-7}^{+8}$ & $2.8_{-0.7}^{+0.8}$ & 0.89 (437/493) \\
C & $5_{-2}^{+5}$ & $0.5_{-0.2}^{+0.3}$ & $104 \pm 10$ & $4.2_{-1.3}^{+6.0}$ &
 $0.2_{-0.1}^{+0.2}$ & $1.0 \pm 0.2$ & $0.3_{-0.2}^{+1.0}$ & $12_{-3}^{+27}$ & $1.6_{-0.7}^{+2.6}$
 & 0.89 (436/491) \\
\hline
\end{tabular}
\end{minipage}
\end{table*}

\begin{figure}
\centering
\resizebox{\hsize}{!}{\includegraphics{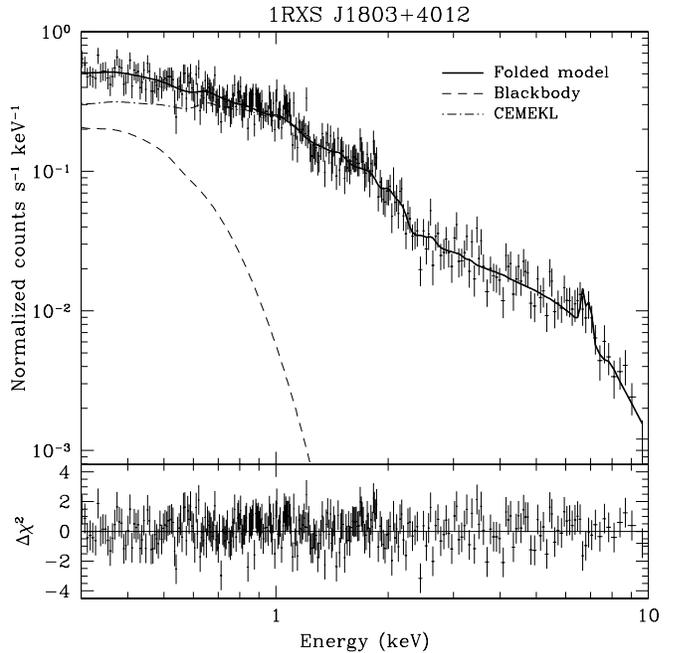}}
\caption{The EPIC-pn spectrum of RXJ1803 fitted with Model B discussed in the text (see
Table~\ref{tab:spectra1803}). The combined MOS spectrum, also used in the fit, is not shown.
Each spectral component that contributes to the best-fit model is shown separately.
The bottom panel reports the residuals expressed in terms of $\sigma$.}
\label{fig:rxj1803_spect}
\end{figure}

The EPIC-pn (Fig.~\ref{fig:rxj1803_spect})
and combined MOS spectra averaged over the whole observation
were fitted using a model constituted by an absorbed MEKAL with solar abundances. This simple model
does not satisfactorily fit the spectrum ($\chi_{\nu}^2 = 2.2$). A slight improvement is achieved by
using a multi-temperature plasma CEMEKL ($\chi_{\nu}^2 = 1.95$) with $k T_{\max} \sim 25\ \mathrm{keV}$
and power-law slope $\alpha$ fixed at unity. In both cases, there is an excess of counts below 0.5~keV
and above 5~keV, but the hydrogen column density of the total absorber drops to zero, thus indicating
that this component is not required. As a comparison, the total galactic absorption in the direction
of RXJ1803 is $N_{\mathrm{H,\,gal}} = 3.4 \times 10^{20}\ \mathrm{cm}^{-2}$ \citep{dic90}. Leaving metal
abundance free, the fit still remains unsatisfactorily high ($\chi_{\nu}^2 = 1.4$) giving sub-solar
abundances $A_\mathrm{Z} = 0.2$. A great improvement ($\chi_{\nu}^2 = 0.91$) of the fit is found
with a composite model (Model A in Table~\ref{tab:spectra1803}) consisting of a blackbody and the
multi-temperature plasma, which decreases the excess of counts at low energies. The blackbody
temperature is found at $k T_{\mathrm{BB}} \sim 100\ \mathrm{eV}$. A further improvement is found by
adding a partial covering (40\%) dense ($N_{\mathrm{H}} = 2.8 \times 10^{23}\ \mathrm{cm}^{-2}$)
absorber (Model B in Table~\ref{tab:spectra1803}). The presence of this component is significant
at the 98.6\% level with the F-test. Here we note that the metal abundance of the optically thin
plasma is still sub-solar ($A_\mathrm{Z} = 0.4$) and that the slope of the power-law emissivity
function was fixed to unity, because when left free it assumed the value $\alpha = 0.67$ but
consistent with unity within errors. A similar quality fit ($\chi_{\nu}^2 = 0.89$, Model C) is obtained
by using two MEKAL components instead of the CEMEKL, supporting the idea that the
optically thin plasma is not isothermal. The parameters of the partial absorber and the blackbody
are consistent with that of Model B, but badly constrained. The temperatures of the two MEKAL components
are 1~keV and 12~keV, the latter being consistent within errors with the maximum temperature of the
multi-temperature plasma found in Model B. The low temperature plasma is instead required to fit
a shallow bump located aroud 1~keV and ascribed to a complex of spectral lines likely dominated by
a strong \ion{O}{VIII} line at 19~\AA. Unfortunately, we cannot inspect the RGS
because of the lack of data from RGS2. The observed X-ray flux in the 0.2--10.0~keV range is
$F_{0.2-10.0\ \mathrm{keV}} = 1.84 \times 10^{-12}\ \mathrm{erg\ cm}^{-2}\ \mathrm{s}^{-1}$.
A comparison with the observation carried out during RASS shows that RXJ1803 was about 1.5 times
brighter in the soft ROSAT band during the XMM-Newton observation. In 2007 the source is 
also less soft than when detected during RASS, since the ROSAT hardness ratios would have been
$\mathrm{HR1} = -0.09$ and $\mathrm{HR2} = +0.12$, suggesting that the soft X-ray component was less
prominent in the XMM-Newton observation than during RASS.

We also performed a phase-resolved analysis of the EPIC-pn spectrum extracted around the pulse
maximum ($\phi = 0.85- 1.15$) and minimum ($\phi = 0.35 - 0.65$), in order to investigate the 
mechanism responsible for the pulsation. Model B shown in Table~\ref{tab:spectra1803} was used,
keeping the metal abundance fixed to the value obtained for the average spectrum and fixing the
power-law index $\alpha = 1$. The parameters of the CEMEKL component are very similar in both fits
and are completely consistent with the value found in the average spectrum. This was expected since
we do not observe any variabilty above 3~keV. Given quoted errors, the main variations
arise in the normalization of the blackbody component, which increases at the spin maximum by a factor
of 2.7 with only a weak evidence that this occurs in the partial absorber column density. The lack of
a significant variability in the hardness ratios suggests that the X-ray spin pulse in
RXJ1803 is likely dominated by changes in the projected area of the blackbody component. It is
interesting to note that the ratio between the spin and the beat amplitudes is larger at lower
energies, which could suggest a much lower contribution of the reprocessed soft X-ray component
to the beat variability.

\begin{table}
\caption{Best-fit spectral parameters of the phase-resolved EPIC-pn spectra of RXJ1803 extracted in
the phase intervals quoted in the text. Model B shown in Table~\ref{tab:spectra1803} has been used.} 
\label{tab:phspec1803}
\centering
\begin{tabular}{lll}
\hline\hline
Parameters 						& Maximum					& Minimum \\
\hline
$F_{0.2-10.0\ \mathrm{keV}}$	& 1.94 						& 1.55 \\
($10^{-12}\ \mathrm{erg\ cm}^{-2}\ \mathrm{s}^{-1}$) &		& \\
\hline
$N_{\mathrm{H}}$ ($10^{23}\ \mathrm{cm}^{-2}$) & $2.9_{-1.6}^{+4.3}$ & $1.4_{-0.9}^{+4.2}$ \\
Cov. Frac.						& $0.5 \pm 0.2$				& $0.6_{-0.2}^{+0.1}$ \\
$k T_{\mathrm{BB}}$ (eV)		& $103_{-17}^{+18}$			& $106_{-45}^{+57}$ \\
$N_{\mathrm{BB}}$ ($10^{-6}$)	& $4.3_{-1.2}^{+3.2}$		& $1.6_{-0.8}^{+1.4}$ \\
$k T_{\max}$					& $14_{-4}^{+6}$			& $13_{-5}^{+15}$ \\
$N_{\mathrm{CEM}}$ ($10^{-3}$)	& $3.4_{-0.6}^{+1.0}$		& $3.2_{-0.6}^{+1.9}$ \\
\hline
$\chi_{\nu}^2$ ($\chi^2$ / d.o.f.) & 1.27 (131/110)			& 0.80 (50/62) \\
\hline
\end{tabular}
\end{table}

\subsubsection {The spectroscopic period of RXJ1803}
                                                                                
\begin{figure}
\centering
\resizebox{\hsize}{!}{\includegraphics{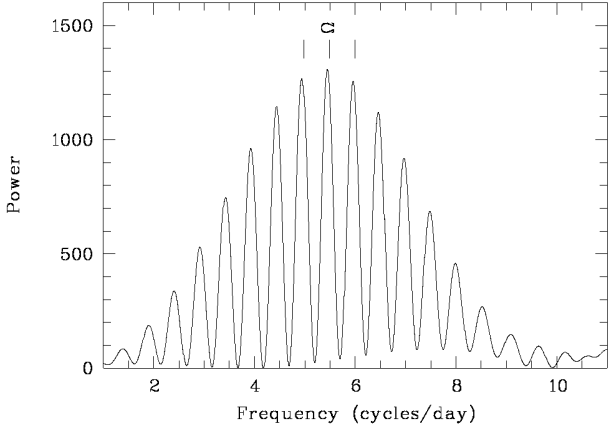}}
\caption{Power spectrum of the $\mathrm{H}_{\alpha}$ radial velocities. The best value is found
at $P_{\Omega} \sim 5.45\ \mathrm{day}^{-1}$ ($\sim 4.40\ \mathrm{hr}$). Secondary maxima are 2
day aliases.}
\label{fig:sp_power}
\end{figure}

The mean optical spectrum of RXJ1803 was found similar to that already reported by \citet{gan05}
and is typical of magnetic CVs with strong emission lines of the Balmer series, \ion{He}{II}
(4686~\AA) and \ion{He}{I} (4471, 5875, 6678 and 7065~\AA), superimposed on a relatively blue
continuum. The radial velocities for the main lines were measured using a single Gaussian least
square fit procedure and periodicities were searched using a Fourier analysis. The power spectrum
for the strongest $\mathrm{H}_{\alpha}$ line gives the most accurate period determination (see
Fig.~\ref{fig:sp_power}). The maximum power is seen at $5.45 \pm 0.27\ \mathrm{day}^{-1}$
with secondary peaks which are 1 and 2~days aliases, corresponding to the night separations.
The characteristics of the radial velocity line modulation were further determined from a
$\chi^{2}$ sine fit. The minimum $\chi^{2}$ corresponds to a period of $P_{\Omega} =
4.402 \pm 0.014\ \mathrm{hr}$, where the error bar is at a $1 \sigma$ level, computed for two
independent parameters. The orbital ephemeris for RXJ1803 is determined as $T_{\Omega} =
\mathrm{HJD}\ 2453917.361(20) + 0.18342(57)\ \mathrm{E}$, where $T_{\Omega}$ is the predicted
heliocentric time of the blue-to-red radial velocity transition.

\begin{table}
\begin{minipage}[t]{\hsize}
\caption{Parameters of the orbital modulation of the emission line radial velocities.
Errors are at the $1 \sigma$ confidence level.}
\label{tab:velocity}
\centering
\renewcommand{\footnoterule}{}
\begin{tabular}{llll}
\hline \hline
Lines &  $\gamma$ ($\mathrm{km\ s}^{-1}$)  & K ($\mathrm{km\ s}^{-1}$) &  Phase~\footnote{Phase of
 the blue-to-red zero crossing}  \\
\hline
$\mathrm{H}_{\alpha}$	& $-94.6 \pm 4.9$   & $69.2 \pm 6.8$    & 0.0  (fixed)   \\  
$\mathrm{H}_{\beta}$	& $-70.3 \pm 6.4$   & $72.2 \pm 8.7$    & $0.02 \pm 0.13$ \\
$\mathrm{H}_{\gamma}$	& $-39.3 \pm 9.1$   & $37.9 \pm 12.6$   & $-0.01 \pm 0.15$  \\ 
\hline
\end{tabular}
\end{minipage}
\end{table}

\begin{figure}
\centering
\resizebox{\hsize}{!}{\includegraphics{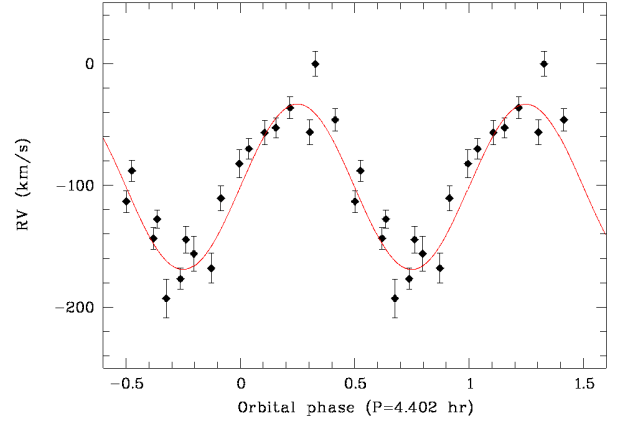}}
\caption{Radial velocities of the $\mathrm{H}_{\alpha}$ line folded with the orbital period
$P_{\Omega} = 4.402\ \mathrm{hr}$. The best sine fit is also shown (line).}
\label{fig:sp_velcor}
\end{figure}

The radial velocities of the $\mathrm{H}_{\alpha}$ line, folded at the above best orbital
ephemeris, are shown in Fig.~\ref{fig:sp_velcor} and the parameters of the corresponding best
sine fits for the strongest emission lines are given in Table~\ref{tab:velocity}. Though the
4.40~hr period is the most significant, we cannot exclude the 2 day aliases at
$4.029 \pm 0.014\ \mathrm{hr}$ and $4.846 \pm 0.014\ \mathrm{hr}$. These observations therefore do
not confirm the period gap value $2.670 \pm 0.008\ \mathrm{hr}$ previously reported by \citet{gan05}.
We note however that a large scatter is visible in their Fig.~22 showing the radial velocities
of $\mathrm{H}_{\alpha}$ wings folded at the 2.67~hr period and that these are
only marginally represented by a sinusoidal
fit at their period. A significant distortion of the line profiles may have been introduced
by the 600~s and 1800~s exposures used, not commensurate with the spin value. Using the 4.40~hr value
for the orbital period and the optical spin value (1520.51~s), the $\omega - \Omega$ and $\omega +
\Omega$ beat periods are predicted at $1690.7 \pm 8.4\ \mathrm{s}$ and $1393.4 \pm 6.9\ \mathrm{s}$,
respectively. The XMM beat value of $1697 \pm 22\ \mathrm{s}$ is therefore consistent with the
above orbital period.

\section{Discussion}

The X-ray analyses of RXJ0704 and RXJ1803 reveal strong pulses at the optical periods previously
identified by \citet{gan05} that now can be interpreted as the true rotational periods of the accreting
WDs and hence confirming these 2 CVs as IPs. This increases the current roster of confirmed IPs to
31 systems~\footnote{For a comprehensive list of IPs see also
\texttt{http://asd.gsfc.nasa.gov/Koji.Mukai/iphome/iphome.html}}. Both sources show pulsations that
are strongly energy dependent in the soft X-ray portion of the spectrum. RXJ0704 and RXJ1803 also
show common spectral properties characterized by a hard and a soft emission component. The latter
property adds these two systems to the small but growing group of `soft IPs'. Now we discuss
their pulse properties and spectral characteristics highlighting their common aspects.

\subsection{RXJ0704} \label{sec:discuss0704} 

The 480~s spin pulsation observed below 2~keV is highly structured with a strong dip feature and a bump,
while no modulation is detectable at higher energies. The spin pulse is very similar to that observed
in PQ~Gem \citep{demartino04,evans06}, which also shows a marked dip that preceeds the spin maximum.
The dip in both sources is due to an increase of absorption from an intervening column along the line
of sight. While for PQ~Gem the geometry of the system and its accretion flow is rather well determined
\citep{potter97}, that in RXJ0704 has still to be established. Notwithstanding this, the strong
similarity of the pulse may suggest that also in RXJ0704 the dip could result from the obscuration of
the magnetic pole produced by the accretion flow along field lines (curtain, \citet{ros88}) which
preceeds the pole itself. In this configuration, the material couples at large distances and travels
out of the orbital plane producing large absorption effects. At later phases, when the
curtain moves away from the line of sight, the X-rays are less absorbed and the WD area of the
accreting pole is still at its maximum visibility. This is also corroborated by the variability seen
at UV wavelengths, where the maximum is observed in correspondence with the X-ray maximum. RXJ0704
has been observed in a previous epoch to show a double-sine optical modulation \citep{gan05}, thus
indicating that both poles are active at some epochs. To have both of them come into view the
magnetic dipole should be relatively inclined ($m \leq i$, where $m$ is the magnetic colatitude
and $i$ is the binary inclination). Eclipses are not observed in RXJ0704, implying $i \la 70^{\circ}$.
Furthermore, the lack of modulation in the hard X-rays suggests that either the accretion shock
does not pass behind the limb of the WD or there are two poles equally contributing to the hard
X-rays. This last is favoured in the framework of two visible poles.
Wether material couples at large radii, producing very localized absorbing structures,
need to be explored with further observations such as phase resolved spectroscopy and multi-band
photometry.

The spectral analysis confirms the presence of a blackbody-like soft X-ray component which strongly
dominates the pulse below 0.5~keV. This component is relatively hot and should originate in a
relatively small area of the WD polar cap. In fact, the normalization constants of the blackbody
in both the epochs of observation imply an emission area that is in the range $4-7 \times
10^{12}\ d_{800\,\mathrm{pc}}^2\ \mathrm{cm}^{2}$~\footnote{For RXJ0704 and RXJ1803 we adopt
lower limits to their distances of 800~pc and 500~pc, respectively. These are obtained using the
2MASS K-band magnitudes, adopting for the secondaries a spectral type  $\mathrm{M3} \pm 2$ for
RXJ0704 and $\mathrm{M3} \pm 1$ for RXJ1803, that are predicted for their orbital periods of 4.17~hr
and of 4.4~hr \citep{knigge06}, and assuming that these donors totally contribute to the K-band
flux.}. The bolometric flux of the blackbody is $F_{\mathrm{BB}} = 2.6 - 5.6 \times
10^{-12}\ \mathrm{erg\ cm}^{-2}\ \mathrm{s}^{-1}$. On the other hand, the optically thin emission
has a bolometric flux 4.6 - 1.4 times larger than that of the soft component in the two observations.
Therefore, to obtain an estimate of the mass accretion rate, we need to include both components. Here
we note that the UV flux is modulated at the spin period by 12\% and, when dereddened by $E_{B-V}
= 0.015$ (as derived from the hydrogen column density of the total absorber, \citet{ryter75}), is
about three orders of magnitude higher than the extrapolated blackbody flux in that range. Hence,
the accretion luminosity is estimated as $L_{\mathrm{accr}} \ga L_{\mathrm{BB}} + L_{\mathrm{hard}}
+ L_{\mathrm{UV}} = 1.1 \times 10^{33}\ d_{800\,\mathrm{pc}}^2\ \mathrm{erg\ s}^{-1}$. The upper
temperature value of the optically thin component of 24~keV provides a lower limit to the shock
temperature, which in turn would give a WD mass of $0.52 M_{\sun}$. We then assume a canonical WD mass
value of $0.6 M_{\sun}$ WD and derive an accretion rate of $\sim
10^{16}\ d_{800\,\mathrm{pc}}^2\ \mathrm{g\ s}^{-1}$. At a distance of 800~pc it is lower by
one order of magnitude than the $2.6 \times 10^{17}\ \mathrm{g\ s}^{-1}$ expected for its orbital
period \citep{warner}.

\citet{norton99} proposed that fast rotators like V709~Cas ($P_{\omega} = 313\ \mathrm{s}$), YY~Dra
($P_{\omega} = 529\ \mathrm{s}$) and V405~Aur ($P_{\omega} = 545\ \mathrm{s}$) possess weakly
magnetized WDs. RXJ0704, with its short 480~s spin period, joins these fast rotators. Its
magnetic moment could be estimated assuming that the WD is accreting from a disc. The strong spin
variability indicates that accretion is disc-fed \citep{norton93}. Hence, the condition for disc
formation and truncation at the magnetospheric radius implies that $R_{\mathrm{mag}} \approx
R_{\mathrm{co}}$, where the corotation radius is the radius at which the magnetic field rotates with
the same Keplerian frequency of the inner edge ot the accretion disc. These are defined as
$R_{\mathrm{co}} = \left( G M_{\mathrm{WD}} P_{\omega}^2 / 4 \pi^2 \right)^{1/3}$ and
$R_{\mathrm{mag}} = 5.5 \times 10^{8} \left( M_{\mathrm{WD}} / M_{\sun}\right)^{1/7} R_{9}^{-2/7}
L_{33}^{-2/7} \mu_{30}^{4/7}\ \mathrm{cm}$, where $R_{9}$ is the WD radius in units of
$10^{9}\ \mathrm{cm}$, $L_{33}$ is the luminosity in units of $10^{33}\ \mathrm{erg\ s}^{-1}$, and
$\mu_{30}$ is the WD magnetic moment in units of $10^{30}\ \mathrm{G\ cm}^3$. For a $0.6 M_{\sun}$
WD, we find $\mu \sim 1.0 \times 10^{32}\ d_{800\,\mathrm{pc}}\ \mathrm{G\ cm}^3$. The spin-to-orbit
period ratio of RXJ0704 is $P_{\omega} / P_{\Omega} \sim 0.032$, thus it is a relatively highly
asynchronous system which poses the question of whether this system is spinning at equilibrium.
Spin equilibria have been computed by \citet{norton04} for systems with a fixed mass ratio
$q = 0.5$ and accreting at the secular value for their orbital periods. For $P_{\omega} / P_{\Omega}
= 0.03$ and an orbital period of 4.2~hr, it is predicted that equilibrium is reached for a magnetic
moment of $\mu_{\mathrm{eq}} \sim 5 \times 10^{32}\ \mathrm{G\ cm}^3$. While this could suggest that
RXJ0704 is not spinning at equilibrium, this issue remains open until the distance and the mass 
ratio remain undetermined.

\subsection{RXJ1803}

The X-ray timing analysis of RXJ1803 has revealed the presence of a strong pulsation, with an amplitude
of 54\%, at the previously identified optical period, which confirms this system as an IP. We
also detect a weaker modulation (by a factor of $\sim 2$) at a period of $1697 \pm 22\ \mathrm{s}$
that we interpret as the beat period. This is consistent with the value obtained using the optical
photometric spin period and our new spectroscopic orbital period, i.~e. $1681.88 \pm 0.56\ \mathrm{s}$.
Hence, both our new spectroscopy and X-ray data show that the binary period is much larger than
previously determined \citep{gan05}. With the new 4.4~hr orbital period, RXJ1803 is outside the
orbital period gap and its spin-to-orbit period ratio lowers to $P_{\omega} / P_{\Omega} \sim 0.096$,
which is well within the range found for most systems \citep[see][]{norton04}.

X-ray pulsations at the beat frequency occur when the accretion stream is directly channelled to the
magnetic poles of the WD, without passing through the disc \citep{wynnking92}. However, the detection
of both spin and beat periodicites suggests an hybrid accretion mode, i.~e. occurring simultaneously
via a disc and disc-overflow \citep{hellier95}. The proportion between the two accretion mechanisms
can be inferred from the relative amplitudes of the corresponding X-ray variabilities. Disc-overflow
was also theoretically predicted by \citet{armitage_livio98}. To date, only the IP V2400~Oph is known
to be a pure stream-fed accretor. A few other systems (such as FO~Aqr, AO~Psc, TX~Col) have shown both
spin and beat variabilities, with different amplitudes. RXJ1803 thus joins these few systems with
disc accretion representing $\sim 66\%$ of the total flow.

The X-ray spin pulse below 3~keV is single peaked and energy dependent, while no detectable
modulation is found at higher energies. Both hardness ratios in the soft bands and spectral analysis
show that the spin pulse is likely dominated by aspect variations rather than photo-electric absorption.
We also find changes of the phase of pulse maximum on a timescale of hours, commensurable with the
orbital period. The X-ray beat variability does not appear to possess the same spectral dependence of the
spin, although it is also observed below 3~keV. The spectrum of RXJ1803 is well described by an
optically thin multi-temperature plasma with maximum temperature $k T_{\max} \sim 30-40\ \mathrm{keV}$
which can be regarded as the shock temperature. A blackbody component at 95~eV also
contributes to the soft portion of the spectrum by $27\%$ of the total observed flux below 1~keV.
Both components are absorbed by a dense ($N_{\mathrm{H}} = 3 \times 10^{23}\ \mathrm{cm}^{-2}$) local
material covering $\sim 40\%$ of the X-ray emitting region. If the latter is mostly related to the
overflowing material, this could explain the different energy dependence of the beat and spin
variabilities. The blackbody emission is also relatively hot, with a very small area on the WD pole
($\sim 1.0 \times 10^{11}\ d_{500\,\mathrm{pc}}^2\ \mathrm{cm}^{2}$). Its bolometric flux is
$F_{\mathrm{BB}} = 2.9 \times 10^{-13}\ \mathrm{erg\ cm}^{-2}\ \mathrm{s}^{-1}$ which results
to be 14 times smaller than that of the optically thin component. To evaluate the accretion luminosity
we include both hard and soft X-ray components, as well as the modulated flux in the UV/optical
bands. However, the UV range does not show a modulation. We therefore include only the B band modulated
flux of $7.3 \times 10^{-14}\ \mathrm{erg\ cm}^{-2}\ \mathrm{s}^{-1}$, corresponding to a luminosity
$2.2 \times 10^{30}\ d_{500\,\mathrm{pc}}^2\ \mathrm{erg\ s}^{-1}$. The accretion luminosity is then
$L_{\mathrm{accr}} \ga L_{\mathrm{hard}} + L_{\mathrm{BB}} + L_{\mathrm{B}} = 1.3 \times
10^{32}\ d_{500\,\mathrm{pc}}^2\ \mathrm{erg\ s}^{-1}$. To estimate the mass accretion rate we use
a WD mass inferred from the shock temperature of 30--40~keV: $M_{\mathrm{WD}} \sim
0.61-0.76\, M_{\sun}$. This gives an accretion rate $\dot{M} \ga 0.8 - 1 \times
10^{15}\ d_{500\,\mathrm{pc}}^2\ \mathrm{g\ s}^{-1}$. If the source is indeed at 500~pc, this value
is low and much smaller than the secular value predicted for its 4.4~hr orbital period, i.~e. $3
\times 10^{17}\ \mathrm{g\ s}^{-1}$ \citep{warner}. Although this is a lower limit to the accretion
rate and unless most of the accretion luminosity is not emitted at high energies, RXJ1803
seems to be a soft IP with a low accretion rate.

The ratio between the spin and orbital period is 0.096, which is very close to the typical value of
0.1 for IPs and suggests that RXJ1803 is spinning close to equilibrium. The dominance of the spin
pulsation also indicates that RXJ1803 accretes predominantly via a disc. As done in
Sect.~\ref{sec:discuss0704}, the condition for disc truncation at $R_{\mathrm{mag}}$ gives a lower
limit for the magnetic moment $\mu \sim 1.3 \times 10^{32}\ d_{500\,\mathrm{pc}}\ \mathrm{G\ cm}^3$
using the $\dot{M}$ and the WD mass derived for this object. Following \citet{norton04}, the magnetic
moment predicted for a system in spin equilibrium with $P_{\omega} / P_{\Omega} = 0.096$ and
$P_{\Omega} = 4.4\ \mathrm{hr}$ is $\mu = 2 \times 10^{33}\ \mathrm{G\ cm}^3$. Further observations
to constrain the multicolour variability as well as the distance of this system are needed to draw
firm conclusions on the mass accretion rate and the magnetic moment of this system.

\subsection{Soft X-ray IPs: an emerging new class?}

\begin{figure}
\centering
\resizebox{\hsize}{!}{\includegraphics{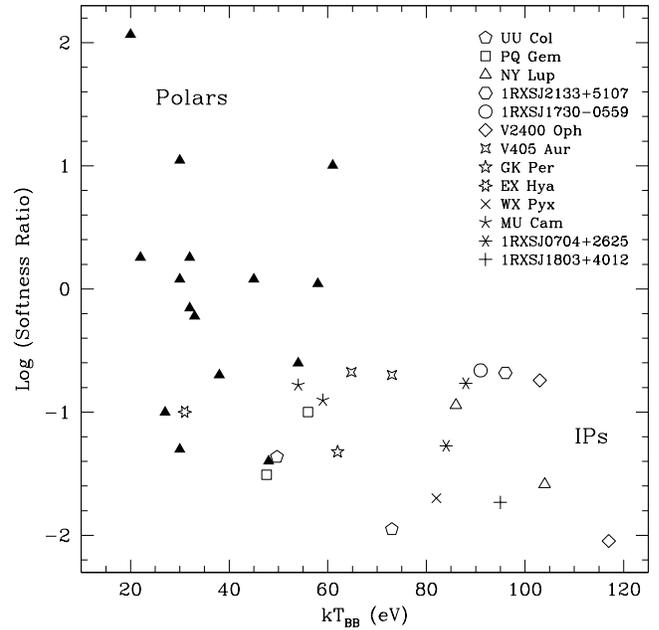}}
\caption{The softness ratio of Polars (filled triangles) and of `soft IPs' (other symbols) versus
blackbody temperature. Values for Polars are taken from \citet{ramsay04}, while those for IPs
from Table~\ref{tab:soft_ips}.}
\label{fig:soft_ratio}
\end{figure}

Our XMM-Newton observations of two new IPs allowed us to detect a soft X-ray blackbody component
in these systems. The ROSAT satellite  was the first to discover three IPs with a significant soft
X-ray component, namely PQ~Gem, V405~Aur and UU~Col
\citep{mason92,haberletal94,motchetal96,burwitzetal96}. Such a component was also recognized in a
few other systems observed with BeppoSAX  (V2400~Oph, \citet{demartino04}) and with the high sensitive
XMM-Newton data (NY~Lup, \citet{haberl02}; 1RXS~J2133.7+5107, \citet{demartino06b}; 1RXS~J173021.5-055933, 
\citet{demartino08}; 1RXS~J062518.2+733433, \citet{staude08}). A systematic search for this
component in the XMM-Newton archival data has further added three systems EX~Hya, GK~Per and WX~Pyx
\citep{evanshellier07}. The current roster of soft X-ray IPs amounts to thirteen systems, which is
now a statistically significative sample, comprising
42\% of all the confirmed IPs. Therefore, `soft
IPs' seem to represent a well defined class among IPs. This makes the soft X-ray emission not a sole
characteristics of Polar systems which deserves an explanation. Moreover, the properties of the
soft X-ray component in IPs still have to be defined. With thirteen systems, this is now possible.
We compiled Table~\ref{tab:soft_ips}, a listing of the blackbody temperatures and softness ratios
(defined as the bolometric flux ratio $F_{\mathrm{soft}} / 4 F_{\mathrm{hard}}$, see \citet{ramsay04})
as derived from spectral analyses available in literature. Despite the great spread due to different
models applied and/or different spectral data, there is clear evidence that the soft X-ray
blackbody in IPs covers a wide range of temperatures (30--100~eV), that is much wider than in
the Polar systems (see also Fig.~\ref{fig:soft_ratio}). In most cases the temperatures are high and
larger than those found in the Polars with only a few systems falling in the same range. These hot
blackbodies could arise from small cores of the accretion spots onto the WD poles. As already noted
in \citet{demartino08}, although the emitting areas are subject of great uncertainties, mostly due to
the distance, systems showing hot blackbodies like NY~Lup, 1RXS~J173021.5-055933, RXJ0704 and RXJ1803
have fractional areas $f \la 10^{-5} - 10^{-6}$ which are two-three orders of magnitude lower than
those found in IPs with cooler blackbodies such as PQ~Gem \citep{jamesetal02}. Since the inferred
blackbody temperature is an average value over the heated spot, the wide range of temperatures might
reflect accretion spots of different sizes.

Strong evidence is found in most Polars that the  soft X-ray component arises in the WD photosphere
primarly due to the heating by accretion of dense blobs
\citep{kuijpers_pringle82,woelk_beuermann96,beuermann99,beuermann04}, with most of the reprocessed
primary (hard X-rays and cyclotron) radiations mainly emitted in the far-UV \citep{konig06}. As seen
in Fig.~\ref{fig:soft_ratio}, the softness ratio of IPs is generally smaller than that observed in
the Polars and, as already noted by \citet{evanshellier07}, this indicates that blobby accretion
does not occur in IPs. We further stress that the presence of a disc in these systems indeed makes
this accretion mode very unlikely. Most IPs also have softness ratios below 0.25, lower than that
predicted by the standard model $L_{\mathrm{repr}} \approx 0.5 \left( L_{\mathrm{brem}} +
L_{\mathrm{cyc}}\right)$ \citep{lamb_masters79}, suggesting that the soft X-ray emission does not
fully account for the whole reprocessing. 

\citet{evanshellier07} propose that the appearance of this component is primarly due to geometric
factors because the cool material in the accretion curtains heavily absorbes the soft X-rays when
viewing along the curtain. They note that those IPs with light curves dominated by absorption dips
tend not to show a blackbody component. While the high density local absorbing material certainly
adds further complications in IPs this might not be the whole story. Our finding of a wide range
of temperatures, and the likely different sizes of the heated spots, could be explained if
bremsstrahlung and cyclotron radiation both play a role in irradiating the WD surface. The fact that
some of the `soft IPs' like PQ~Gem \citep{mason97}, V405~Aur \citep{shakhovskoj_kolesnikov97},
V2400~Oph \citep{buckleyetal95} and 1RXS~J2133.7+5107  \citep{katajainen07}, were found to show
optical/near-IR circular polarization, suggests that cyclotron radiation cannot be neglected. Here
we note that it cannot be excluded that other `soft IPs' are also circularly polarized but have
not been searched in polarized light yet.

Beamed cyclotron radiation originates in the higher parts of the post-shock region and, thus, can
heat large spots. Bremsstrahlung is instead irradiated isotropically from the dense post-shock regions
closer to the WD surface and hence irradiates smaller areas. Therefore we might expect that for higher
magnetic field IPs, cyclotron cooling is not totally negligible and therefore these systems would have
larger accretion spots with respect to other IPs with lower magnetic fields. This in turn would imply
that higher field IPs are expected to show on average lower blackbody temperatures than the low field
IPs. A test to this hypothesis would then require a systematic search for polarized radiation in the
soft IPs and, in particular, over a wide range of wavelengths from visual to IR. 

The detailed and recent study on reprocessed radiation in Polars by \citet{konig06} relies on the clear
evidence of a large irradiated spot on the WD surface in AM~Her that mainly emits in the far-UV,
implying that most of the soft X-rays do not enter in the energy balance. For IPs, there is still a
great uncertainty in whether the UV modulated flux comes from the irradiated WD pole or from the inner
regions of the curtain or both (e.~g. \citet{demartino99,demartino06a,demartino08,eisenbart02,evans06}).
Detailed far-UV/UV coverage to identify the heated WD surface in IPs would be highly desirable to assess
the energy balance in these systems.

\begin{table}
\caption{Blackbody temperatures and softness ratios (see text) of 13 `soft IPs', as
derived from spectral fits available in literature. For some objects different values are reported
as different models were applied and/or different spectral data were used. References are (1)
\citet{demartino06a}, (2) \citet{evanshellier07}, (3) \citet{demartino04}, (4) \citet{haberl02},
(5) \citet{demartino06b}, (6) \citet{demartino08}, (7) \citet{staude08}.} 
\label{tab:soft_ips}
\centering
\begin{tabular}{llll}
\hline\hline
Object	& $k T_{\mathrm{BB}}$ & Softness Ratio & Reference \\
		& (eV)	& $F_{\mathrm{soft}} / 4 F_{\mathrm{hard}}$ & \\
\hline
UU Col				& 50	& 0.043	& 1 \\
					& 73	& 0.011	& 2 \\
PQ Gem				& 56	& 0.1	& 3 \\
					& 48	& 0.031	& 2 \\
NY Lup				& 86	& 0.114	& 4 \\
					& 104	& 0.026	& 2 \\
1RXS J2133+5107		& 96	& 0.208 & 5 \\
1RXS J1730-0559		& 91	& 0.218	& 6 \\
V2400 Oph			& 103	& 0.181	& 3 \\
					& 117	& 0.009	& 2 \\
V405 Aur			& 73	& 0.2	& 3 \\
					& 65	& 0.211	& 2 \\
GK Per				& 62	& 0.048	& 2 \\
EX Hya				& 31	& 0.101	& 2 \\
WX Pyx				& 82	& 0.02	& 2	\\
MU Cam (2005)           & 59    & 0.125 & 7 \\
MU Cam (2006)           & 54    & 0.166 & 7 \\
RXJ0704 (Oct. 2006)	& 84	& 0.053	& This work \\
RXJ0704 (Mar. 2007)	& 88	& 0.172 & This work	\\
RXJ1803				& 95	& 0.019	& This work \\
\hline
\end{tabular}
\end{table}

\section{Conclusions}

XMM-Newton observation of RXJ0704 and RXJ1803 confirm these CVs as true members of the IP class.
Both systems are dominated by strong pulsations at the WD rotational periods which dominate the
soft portion of their spectrum, i.~e. below 3-5~keV. With $P_{\omega} = 480\ \mathrm{s}$, RXJ0704
joins the group of fast rotators and possess a relatively high degree of asynchronism ($P_{\omega}
/ P_{\Omega} = 0.04$). RXJ1803 is, instead, a slow rotator $P_{\omega} = 1520.5\ \mathrm{s}$. From
our new optical spectroscopy and the X-ray data of this system we infer an orbital period of
4.4~hr, that is longer than previously determined. This gives a spin-to-orbit period ratio of
$\sim 0.1$, typical for IPs.

The detection of strong spin pulses indicate that accretion occurs preferentially via a disc in
both systems. However, while RXJ0704 only shows X-ray variability at the spin period, RXJ1803 also
displays a weak variability at the orbital sideband ($\omega -\Omega$), implying that accretion
also occurs via disc-overflow.

Their X-ray spectra are compatible with composite emission consisting of an optically thin component
($\sim 11-25\ \mathrm{keV}$), which in RXJ1803 is better described by a multi-temperature plasma,
reaching values up to 30--40~keV, as well as by blackbody emission with a relatively high temperature
(85--100~eV) heavily absorbed by dense material partially covering the X-ray source. This adds these
two systems to the growing group of `soft IPs', which indicates that the presence of a soft blackbody
component is not solely a characteristic of the Polars. However, a major difference is in the
temperatures of the blackbodies in the IPs which cover a wider range than that observed in the Polar
systems. We qualitatively try to explain the differences in terms of reprocessing over different sizes
of the WD spot. We suggest that IPs showing cooler soft X-ray blackbody components also possess WDs
irradiated by cyclotron radiation. A further systematic observational approach addressing the detection
of polarized optical/IR emission in IPs is highly desiderable.

\begin{acknowledgements}
We acknowledge the XMM-Newton SOC and MSSL staff for help in the reprocessing of the OM data.
Prof.~A.~Bianchini is gratefully acknowledged for useful discussion. DdM and GA acknowledge financial
support from ASI under contract ASI/INAF I/023/05/06 and from INAF under contract PRIN-INAF 2007 N.17.
\end{acknowledgements}

\end{document}